# CURRENT SERVICES IN CLOUD COMPUTING: A SURVEY

Mohamed Magdy Mosbah[1], Hany Soliman [2], Mohamad Abou El-Nasr [3]

[1] Department of Information Systems, Arab Academy for Science, Technology & Maritime Transport, Cairo, Egypt
[2] Department of Computer Engineering, Cairo higher institute for engineering, computer science & management, Cairo, Egypt
[3] Department of Computer Engineering, Arab Academy for Science, Technology & Maritime Transport, Alexandria, Egypt

## ABSTRACT

*Due to the fast development of the Cloud Computing technologies, the rapid increase of cloud services are became very remarkable. The fact of integration of these services with many of the modern enterprises cannot be ignored. Microsoft, Google, Amazon, SalesForce.com and the other leading IT companies are entered the field of developing these services. This paper presents a comprehensive survey of current cloud services, which are divided into eleven categories. Also the most famous providers for these services are listed. Finally, the Deployment Models of Cloud Computing are mentioned and briefly discussed.*

## 1. INTRODUCTION

A cloud computing (Cc) is a technology that has the fastest growth rate in the field of information technology (IT) and showed a high growth rate during the last few years. Cc is currently one of the biggest buzz words and the amount of Cc services is increasing quickly. Many companies that could be considered as the giants of software industry like Microsoft and other heavyweight companies in the field of Internet technology, such as Amazon and Google are joining to develop cloud services [1], [2]. Cc provides a brand new standard for service provisions from the beginning, with low upfront investment, expected performance, infinite scalability, high availability, and so on.

A new standard for the provision of services in the cloud computing include three roles [3]. As shown in Figure 1, according to the service Patterns, they cloud be divided into eleven Categories, and offered through four different deployment models and have been provided by several of current leading companies

The aim of this paper is to provide a clear view for the current services of the cloud, which are divided into eleven categories, companies leading of cloud computing services, and deployment models.

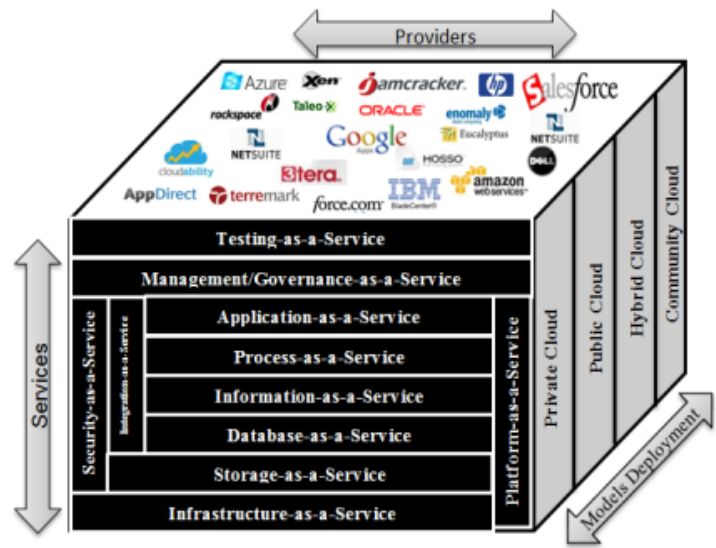

Figure 1. The Relationship between Services, Deployment Models, and Providers

The rest of this paper is organized as follows: In section 2, the paper presents the current eleven categories of Cc services. In section 3, the four deployment models of Cc are explained. Finally, Concluding remarks in the last of the paper.

## 2. CURRENT CLOUD COMPUTING SERVICES

### 2.1. Systems on Storage as Services

Storage as a Service (StaaS) allows users to remotely store information and enjoy the on-demand high quality cloud applications without the workload local hardware and software management. An amount of leading companies are providing storage as a service such as DropBox, AT&T and so on.

- **DropBox:** DropBox [4] authorize users to create a certain folder on each of their computers, which Drop Box is a free service that grant you the ability to bring your contents e.g. (photos, docs, and videos) anywhere and share them easily. Any file you save to your Drop Box is accessible from all your computers, iPhone, iPad, smart device and even the Drop Box website. The offers start from 2GB of Free space up to 16GB while 2 GB is a free offer. It competes with many companies that offer the same service, such as Google Drive [5], Box.net [6], SkyDrive [7], Amazon Cloud Drive [8] and many other similar services.
- **AT&T:** AT&T synaptic storage as a service [9] provides cloud-based virtual data storage with on-demand scalability, including managed servers, security, and storage. The service is charged in a utility computing manner, pay-as-you go pricing. AT&T can provide easy online sign-up, access to service in a few minutes, and backup to supplement existing data storage services.

## 2.2. Systems on Database as Services

Database as a Service (DbaaS) allow users to access enterprise-grade database over the internet. Database as a service provides the ability to authorize the services of a remotely hosted database. Consume can run database any time without buying hardware, software, and the maintenance activities required for a database. An amount of leading companies are providing DbaaS such as Xeround, Microsoft azure, Amazon RDS and so on. However its strength in exploring the technology database that would be so expensive for providing such specification.

- **Xeround:** Xeround [10] provides a cloud database software service for applications based on the open source version of the MySQL database. Xeround offers its service on several cloud platforms, it supported Amazon EC2 [8], and Rackspace [11], and is planning to support additional providers. The service offers pay-as-you go pricing, calculated per Gigabyte per hour, with an extra charge for data transfer for huge databases.
- **Amazon Web Service (AWS):** Offer a new option to present this need with Amazon Relational Database Service. (Amazon RDS) [8] provides the flexibility of being able to scale the compute resources or storage capacity associated with a relational database instance through a few clicks. Amazon RDS service is charged in a utility computing manner, pay- per-use. The provision of such hardware and install the suitable software version to run database in cloud is unneeded.

## 2.3. Systems on Information as Services

Information as a Service (InaaS) allows any application to access any type of information using API, and refers to the ability to consume any type of remotely hosted information (e.g. address validation, and stock price information). Therefore, as you must for database as a service, you need to consider the interface offered by information as a service. Enterprises use information from many different sources through a single application or mashup [12].

## 2.4. Systems on Process as Services

Sometimes referred to as business process as a service (BPaaS), is a resource that can remote and connect many resources together, such as service and data, either hosted within the same cloud computing resource or remotely to create business processes. Any business process (e.g. ecommerce, payroll and printing) delivered as a service over the internet and accessible from all your computers, Phone, and smart device, can be considered as a (BPaaS). Advertising services that offered by some company such as Google AdSense [13], IBM Blue works Live [14] and so on. According to IBM the BPaaS sits above software as a service layer [15].

## 2.5. Systems on Application as Services

Also known as Software as a Service (SaaS) is any application delivered through the web's platform to an end user, typically as taking the advantage of the application through a browser. Application as a service is the ability to leverage an enterprise-class application without buying and installing enterprise software Application as a service is a model of software publishing according to the provider licenses however the customers can use this service upon their demand, such as Sales force SFA, office automation applications are indeed (AaaS) as well, including Google Docs, and Gmail. An amount of leading companies are providing SaaS, such as Media Temple, 3Tera, and AppNexus, and so on.

- **MediaTemple:** media Temple's Grid-service [16] is a web hosting service. It has eliminated barriers and one of the failures by using hundreds of servers working alongside for customers' sites, applications, and email. The on demand scalability means its customers are always able to make the system work for intense bursts of traffic and growing audience.
- **3Tera:** AppLogic [17] is the first grid operating system service in the world. AppLogic provides many services. It is an integrated system that converts arrays of servers into virtualized resource pools that customers can subscribe to in order to improve the power of their applications. AppLogic makes it easy to navigate between the current web applications without modifications onto a grid. Not only the virtual machines can Customers defined but also complex application infrastructure such as VPNs, firewalls, and storage, all with nothing more than a browser.
- **AppNexus:** AppNexus cloud [18] provides many kinds of services and is very easy to use. Its customers only need to take just 15 minutes to reserve as many servers as they need to use. By using services provided by AppNexus, its customers don't need to long time, cost, and headaches to build an infrastructure, while also avoiding planning the infrastructure investment far ahead. All the services can be adjusted at running time.

## 2.6. Systems on Platform as Services

Platform as a Service (PaaS) is paradigm for delivering operating systems and services over the internet without cost, software download or installation for developers. Platform service including application development, interface development, database development, and storage delivered through a remotely hosted platform to subscribers. An a mount of leading companies are providing cloud platforms, such as Google App Engine, Microsoft Azure, Engine Yard, and Force.com ( part of Salesforce.com) and so on .

- **Google App Engine:** Google App Engine [19] lets customers run their web applications on Google infrastructure. Google App Engine applications are easy to build, data storage, and maintain. And allow to customers Uploading applications to App Engine and starting to serve, no servers are needed to maintain.
- **Microsoft Azure:** Microsoft azure services platform [20] all Azure services and applications built on top of Windows Azure platform. Azure is running on Windows Server 2008 and has a web based configuration site which enable to over the internet. Azure can host applications and run at Microsoft data centers. It serves as a runtime environment for the applications and it also provides a set of services that enables users to manage, develop and host their own applications off premises.
- **Engine Yard:** Engine Yard [21] is a traditional hosting company in the cloud computing literature. It runs small-scale web applications and large-scale enterprise applications. Engine Yard uses security personal to physically control and to monitor access to user's data center for 24/7. Also it takes place between such platforms as a service (PaaS) for a wide range of Ruby on Rails, PHP and Node.js applications.
- **Force.com:** Force.com [22] is one of the platforms which specialized in creating and publishing applications for the social enterprise. However, there are no servers or software neither to buy nor to manage, but you can focus only on creating applications that include built-in social and mobile functionality, business processes, reporting, and search. Applications run on a secure, proven service to grow a business, and backs up data automatically. And allowing developers to build multi-tenant applications (e.g., multi-tenant database) that are hosted on saleforce's servers as a service.

## 2.7. Systems on Integration as Services

Integration as a Service (InaaS) is ability to deliver a complete integration package from the cloud, including interfacing with applications and integration design. InaaS allows consumer to develop, maintain, and manage custom integrations for different systems and applications in the cloud. An amount of leading companies are providing Integration as a service such as Dell Boomi, CloudSwitch, and MuleSoft and so on.

- **Dell Boomi:** Dell Boomi [23] is powered by Atmosphere technology. Boomi is a consolidation system that was built in the cloud to utilize the full value of the cloud like most cloud integration systems; it allows consumers to connect any group of cloud, SaaS or on-premises applications without appliances, no software, and no coding.
- **CloudSwitch:** CloudSwitch [24] integration services enables enterprises to run applications in the right cloud without re-architecting the application or the data center tools, policies, and can navigate smoothly between various cloud environments and back into the data center based on the requirements of the business.
- **MuleSoft:** Mule Soft [25] provides a number of consolidation products that bind both SaaS and on-premise applications together and is delivered as a packaged integration experience. It allows administrators to remote workflows across different applications automatically, both on-premises and in the cloud. It helps solve common cloud-to-cloud and cloud-to- premise application integration problems.

## 2.8. Systems on Security as Services

Security as a Service (SeaaS) is the ability to deliver core security service remotely over the internet. While the typical security services provided primary more developed services such as identity management are becoming available. An amount of leading companies are providing security as a service such as AppRiver, and McAfee, and so on.

- **AppRiver:** AppRiver [26] does messaging security in the cloud. It offers SeaaS for e-mail and Web which is based on security tools that are subscription-based and include spam and virus protection, e-mail encryption and Web security. It also offers a full managed service.
- **McAfee:** McAfee [27] Cloud Security helps enterprise safely and confidently leverage secure cloud computing services and solutions instead of adopting the unique and unknown security practices and policies of each cloud vendor, McAfee cloud security allows businesses to extend and apply their own access and security policies into the cloud by securing all the data traffic moving between the enterprise and the cloud, as well as data being stored in the cloud. It competes with many companies that offer the same service, such as Panda [28], Symantec [29], and many other similar services.

## 2.9. Systems on Management/Governance as Services

Management/Governance as a Service (MGaaS) is a general term for provides the ability to manage one or more cloud services. Governance systems, such as the ability to enforce defined policies on data and services, are becoming available as well. The cloud services governance aims to secure applications and data when they are located remotely. Most enterprises like to control management and governance [30].

## 2.10. Systems on Testing as Services

Testing as a Service (TaaS) gives the ability to test local or cloud delivered systems by using remotely hosted testing software, hardware, services and have the ability to test other cloud applications, web sites, and systems of the internal enterprise, and they do not need a hardware or software. An amount of leading companies are providing testing as a service such as SOASTA, and PushToTest, and so on.

- **SOASTA:** SOASTA [31] is a provider of cloud-based testing services. Website tests include software performance testing, load testing, functional testing and user interface testing. SOASTA provides cloud website testing with their product Cloud Test, which simulates multi-users visiting a website at the same moment. SOASTA allows customers to use tests or create customized tests which defined previously to automatically test their web applications.
- **PushTOTest:** PushToTest [32] Methodology into your hands for a test tool that installs on your desktop runs tests in grid and cloud computing environments, or both. PushToTest provides specific control to support automatic publishing and operation of your tests in a grid of servers and a Cloud Testing environment. e.g., identify a cloud testing service such as Amazon EC2 in a Test scenario. PushToTest Cloud Testing allows you to pay when you operate tests.

## 2.11. Systems on Infrastructure as Services

Infrastructure Infrastructure as a service (IaaS) is the last level of abstraction on the cloud. Allow users to manage, storage, networks, processing and other major computing resources so that they can publish and run software, which can include operating systems and/or applications. The user does not manage or control the hardware cloud infrastructure but has control over operating environments, deployed applications, and storage and possibly select networking components. An amount of leading companies are providing Infrastructure as a service such as, Mosso, Skytap, and so on.

- **Mosso:** Mosso [33] is a division of Rackspace, which is well known for its reliable dedicated web hosting across the world. Its services are called cloud sits. In addition, it also provides cloud file for hosting service and cloud server for computation service.
- **Skytap:** Skytap [34] offers a virtual lab service which provides the capability in setting up accessing and managing customers' own applications which are hosted in the virtual machine in the Cloud like the industry leading hypervisors including Xen, VMware and Microsoft Hyper-V. Skytap supports all operating systems that run on these hypervisors including Solaris, Microsoft Windows and Linux variants. It providers a large set of tools to help customers to develop applications in the virtual lab.

## 3. CLOUD DEPLOYMENT MODELS

The Cloud is consisting of four models which based on their infrastructure [35].

### 3.1. Private Cloud

Only single organization has the cloud infrastructure that has the ability to rent or operate it for its own benefits. No public access to it is permitted.

### 3.2. Public Cloud

The cloud infrastructure is owned by an organization selling cloud service is made available to a large industry group or general public.

### 3.3. Hybrid Cloud

The case of combining of more than one cloud infrastructure (private, community, or public) that remain unique entities but are bound together by standardized or proprietary technology that enables data and application portability (e.g., cloud bursting for load-balancing between clouds).

### 3.4. Community Cloud

Only limited number of organizations can shared cloud infrastructure and supports a specific Community which has shared concerns (e.g. security requirement, policy, and goals).

## 4. CONCLUSION

Cloud computing is based on the demand access to virtualized IT resources that are housed outside of your range , while you can share it with different services however use it easily  in addition you can subscribe for this service with an inexpensive monthly fees, and navigate the web smoothly that has many features. This paper outlined a survey in cloud computing services, focusing on the long list services provided by leading companies. The researchers still have more work to do; we hope this paper will be considered as a starting point identifying opportunities for future research.

International Journal of Computer Science, Engineering and Information Technology (IJCSEIT), Vol.3,No.5,October 2013

8